\documentclass[12pt]{article}
\usepackage{geometry}                		
\usepackage{graphicx}				
\usepackage{amssymb}
\usepackage{subcaption}
\usepackage{bm}
\usepackage{amsmath}
\usepackage{authblk}
\usepackage{caption}
\usepackage{wrapfig}
\usepackage{t1enc}
\usepackage[english]{babel}
\usepackage[latin2]{inputenc}
\usepackage{graphicx}
\usepackage{babel,a4wide,amsfonts,bm}
\usepackage{slashed}
\usepackage{dsfont}
\usepackage{color}
\usepackage{hyperref}
\usepackage{cleveref}

\newcommand{\be}{\begin{eqnarray}}
\newcommand{\ee}{\end{eqnarray}}

\newcommand{\bea}{\begin{align}}
\newcommand{\eea}{\end{align}}

\newcommand{\CK}{{\mathcal{K}}}

\def\bea{\begin{eqnarray}}
\def\eea{\end{eqnarray}}

\def\funp{{I\!\!P}}

\graphicspath{{figs/}}

\usepackage{color}
\usepackage[normalem]{ulem}

\definecolor{ascolor}{rgb}{1,0,1}
\definecolor{wlcolor}{rgb}{0,0.6,0.2}
\definecolor{rscolor}{rgb}{0,0,1}
\definecolor{kkcolor}{rgb}{1,0,0}

\DeclareRobustCommand\asout{\bgroup\markoverwith{\color{ascolor}{\rule[0.4ex]{2pt}{0.8pt}}}\ULon}
\DeclareRobustCommand\wlout{\bgroup\markoverwith{\color{wlcolor}{\rule[0.4ex]{2pt}{0.8pt}}}\ULon}

\DeclareRobustCommand\kkout{\bgroup\markoverwith{\color{kkkcolor}{\rule[0.4ex]{2pt}{0.8pt}}}\ULon}

\title{Revisiting gluon density from the BK equation with kinematical constraint and large $x$ terms}
\author[1]{Krzysztof Kutak}
\author[2]{Wanchen Li\thanks{\texttt{wanchenli@fudan.edu.cn}}}
\author[3]{Anna Stasto}
\author[4]{Robert Straka}

\affil[1]{\normalsize Institute of Nuclear Physics, Polish Academy of Sciences, 
 ul.~Radzikowskiego 152, 31-342, Krak\'ow, Poland}
\affil[2]{\normalsize Department of Physics and Center for Field Theory and Particle Physics, Fudan University, Shanghai 200438, China}
\affil[3]{\normalsize Department of Physics, Penn State University, University Park, PA 16802, U.S.A. }
\affil[4]{\normalsize AGH University of Science and Technology, Krakow, Poland}

\begin{document}
\maketitle

\begin{abstract}
We perform analysis of the small $x$ non-linear evolution equation formulated in momentum space supplemented by higher order terms. The equation is defined in wide range of transverse momentum and longitudinal momentum fraction extending previous studies performed in \cite{Kutak:2003bd,Kutak:2004ym}.
The linear part of the equation is motivated by the renormalization group improved small $x$  approach which accounts for resummation of higher orders, and includes collinear splitting function and kinematical constraint. 
The solution to the equation is then used to  perform the fit to Deep Inelastic Scattering reduced cross section data.
   
\end{abstract}

\section{Introduction}
Present day high energy accelerators, like Large Hadron Collider (LHC) allow for the exploration of the new kinematic regime in particle interactions.  Tests of the Standard Model are thus  possible with unprecedented accuracy in the region where the total center-of-mass energy is much higher than the masses of the incoming hadrons. For the precise description of the processes in these collisions, a detailed knowledge of the partonic structure of the incoming hadrons is necessary.  The information is encoded in the {\em parton distribution functions} (PDFs), which in the usual collinear factorization framework are obtained from the solution to the Dokshitzer-Gribov-Lipatov-Altarelli-Parisi (DGLAP) equations \cite{Dokshitzer:1977sg,Gribov:1972ri,Gribov:1972rt,Altarelli:1977zs}. 
These equations allow for the evolution of {\em integrated} PDFs with some hard scale $Q^2$ and resum the large logarithms $\ln Q^2/Q_0^2$ where $Q_0$ is some reference scale.  The splitting functions for DGLAP evolution are known up to NNLO \cite{Vogt:2004mw,Moch:2004pa} and recent intense theoretical efforts aim push the accuracy beyond that level. 

An alternative framework for the calculation of the processes in the high energy limit is given by the  $k_T$ or {\em high energy factorization}  \cite{Catani:1990eg,Catani:1990xk}, where the structure of the hadron is encoded in the {\em unintegrated} gluon distribution function. This is usually referred to as the high-energy limit or the Regge limit, with $s \gg |t|, m^2$, where $s$ is the center-of-mass energy squared, $t$ is the momentum transfer squared, and $m$ typical masses of the produced particles. The corresponding evolution is governed by the Balitsky-Fadin-Kuraev-Lipatov (BFKL) evolution equation \cite{Balitsky:1978ic,Kuraev:1977fs,Lipatov:1985uk}, which sums up the powers of  logarithms $\ln s/s_0$ where  $s_0$ is a reference scale. 

For the case of the Deep Inelastic Scattering process (DIS) of scattering electrons off protons, the energy scale $s_0$ is taken to be equal to the (negative) virtuality of the photon $Q^2$. Therefore the BFKL resums large logarithms of $\ln 1/x$, where $x\simeq Q^2/W^2$ is the Bjorken $x$ variable with $W^2$ being the square of the photon-proton energy. The solution to the BFKL equation results in the  power like growth of the gluon density  at small $x$ and thus of the corresponding cross sections.
The BFKL evolution equation is known up to NLL \cite{Fadin:1998py,Ciafaloni:1998gs} accuracy in QCD, and up to NNLL level in N=4 sYM theory \cite{Velizhanin:2015xsa,Gromov:2015vua,caron2018high}. The LL solution  gives the power-like growth of the gluon density with the decreasing $x$ as $x^{-\omega_{P}}$, with the LL intercept $\omega_P=4 \ln 2 \frac{\alpha_s N_c}{\pi} \sim 0.5$. The solution at LL leads to the  growth of structure functions that is too steep for the experimental data at HERA, e.g. \cite{Bojak:1996nr}. The NLL order corrections turned out to be large and negative leading to instabilities, and thus triggered the need for the resummation of the higher order terms in order to stabilize the results. 

The major ingredients of the resummation we implement in this paper are the kinematical constraint, DGLAP splitting function and the running of the strong coupling. One of the first approaches, by Kwieci\'nski-Martin-Sta\'sto (KMS)  \cite{Kwiecinski:1997ee},  used the  unified the BFKL and DGLAP evolution. In this work the unified system of equations for the unintegrated gluon density was constructed which was based on the BFKL evolution equation supplemented with the kinematical constraint and the non-singular DGLAP evolution terms. Coupled with the equation for quarks, this system was shown to produce the gluon distribution which could be fitted to the structure function data from HERA,  with only two free parameters. It  contained both $Q^2$ and $x$ evolutions on equal footing. Based on this system of unified equation, a resummation framework was developed by Ciafaloni-Colferai-Salam-Sta\'sto (CCSS) in a series of works \cite{Ciafaloni:1999au,Ciafaloni:1999yw,Ciafaloni:2003ek,Ciafaloni:2003kd,Ciafaloni:2003rd,Ciafaloni:2007gf}. In that approach the Kwieci\'nski-Martin-Sta\'sto (KMS) \cite{Kwiecinski:1997ee} formulation was supplemented by the NLL terms of the BFKL, with appropriate subtractions  as well as the anti-collinear terms. Recently the CCSS resummation was applied to the structure functions at HERA, and good description of the data at small values of $x$ was obtained \cite{Li:2022avs}. 

In the region of very small $x$, there are additional effects that need to be taken into account.  When the gluon density becomes very high, recombination of gluons, will start to play an important role, leading to the taming of the rapid growth. This is referred to as {\em parton saturation} \cite{Gribov:1984tu, Mueller:1985wy}. The effective theory which describes the  limit of high energy and density QCD is the Color Glass Condensate (CGC) \cite{McLerran:1993ka,McLerran:1993ni,McLerran:1994vd}, with the corresponding Jalilian-Marian-Iancu-Weigert-Leonidov-Kovner (JIMWLK) evolution equations \cite{JalilianMarian:1996xn,JalilianMarian:1997dw,JalilianMarian:1997gr,JalilianMarian:1997jx,Iancu:2000hn,Iancu:2001ad}. An equivalent approach is based on the operator expansion at high energy, together with the corresponding Balitsky hierarchy equations \cite{Balitsky:1995ub,Balitsky:1998ya}.  When the large color limit $N_C$ is taken, this hierarchy allows for the decoupling of the nonlinear equation, the Balitsky-Kovchegov (BK) 
 evolution equation \cite{Balitsky:1998ya,Kovchegov:1999ua,Kovchegov:1999yj} for the dipole scattering amplitude. It was later shown that the subleading-$N_C$ corrections is very small, resulting only 0.1\% difference between the JIMWLK and BK simulations \cite{Kovchegov:2008mk}. The BK equation contains the non-linear term responsible for the gluon recombination, and the solution can be characterized by the presence of the dynamically generated, energy dependent,  saturation scale $Q_s(x)$, which divides the region between the dilute and dense regimes. 

The BK equation is originally formulated for the dipole amplitude and thus it is written  in the coordinate space \cite{Balitsky:1995ub,Kovchegov:1999ua,Kovchegov:1999yj}. Recently it has been generalized and solved at NLO accuracy \cite{Balitsky:2007feb,Lappi:2015fma}.  In papers \cite{Kutak:2003bd,Bartels:1994jj,Nikolaev:2006za} the BK evolution equation has been rederived in the momentum space, and in \cite{Kutak:2012rf} this non-linear term was included in the KMS \cite{Kwiecinski:1997ee} formulation. The gluon density obtained by solving this equation  was used to successfully describe the $F_2$ structure function data and was used in numerous phenomenological applications \cite{vanHameren:2014ala,vanHameren:2016ftb,Albacete:2018ruq,Bhattacharya:2016jce,vanHameren:2023oiq}. One assumption of \cite{Kwiecinski:1997ee,Kutak:2012rf} was that the solution was truncated by an infrared cutoff and extrapolated `by hand' to lower transverse momentum $k_T$. On the other hand, it is customary to solve the BK evolution equation for large dipole sizes, when using its solutions to fit to the structure function data \cite{Kutak:2012rf},  where the non-perturbative contribution is usually taken automatically into account by integration over the large dipole sizes with the flat dipole cross section, see e.g. \cite{Berger:2011ew,Mantysaari:2018zdd,Albacete:2010sy, Cepila:2018faq}

In this paper we extend the Kutak-Sapeta (KS) approach of \cite{Kutak:2012rf} and solve the non-linear BK equation in momentum space including running coupling, non-singular part of the DGLAP splitting function for both collinear and anti-collinear regions as  well as kinematical constraint. The non-linear term regulates the behavior at very low transverse momenta, and therefore it allows for extending the solution to this region.
Unlike other approaches for the resummed BK \cite{Iancu:2015joa}, this formulation in momentum space allows for the inclusion of the full DGLAP gluon splitting function in the evolution on an equal footing to the BFKL kernel. 
The initial condition is given by a form motivated by the momentum space Golec-Biernat, Wusthoff (GBW) model \cite{Golec-Biernat:1998zce}, and the fit is performed to the  reduced cross section data from HERA \cite{H1:2015ubc}, and consistent description is achieved. This formulation can be easily extended to include the full CCSS resummation for the linear term, as it has been done in \cite{Li:2022avs}, and also for the case of nuclei, by rescaling the radius parameter by $A^{1/3}$ for the non-linear term. 
Considering the numerical aspects: we iteratively solve an integro-differential evolution equation \cref{eq:virtual_kms_F} for consecutive values of $x$ starting from $x=1$. Our numerical algorithm is implemented in CUDA C that enabled parallel evaluation of the equation \cref{eq:virtual_kms_F} with respect to $k^2$. Moreover, the converged part of solution for previous values of $x$ is used in integration with respect to next value of $x$, parts of DGLAP integrals are precomputed and tabulated. Aforementioned optimizations allowed for a considerable speedup compared to a sequential code.

The structure of the paper is as follows:  In \cref{sec:bk}, we begin with the LL BK equation in momentum space and outline the higher-order resummation improvements implemented in the equation. \Cref{sec:structurefunction} introduces the $k_T$  factorization framework, which allows us to compute the structure functions. In \cref{sec:fitresults}, we present our fits to the HERA data and analyze the behavior of the resulting gluon density. Finally, we conclude in \cref{sec:conclusions}.

\section{LL BK in momentum space equation and resummations}
\label{sec:bk}
\subsection{LL BK equation in momentum space}

The BK equation was originally derived in the coordinate space \cite{Balitsky:1998ya,Kovchegov:1999ua,Kovchegov:1999yj}. It can be reformulated in the momentum space \cite{Kutak:2003bd,Bartels:2007dm}. The equation for  unintegrated gluon density ${\cal F}(x,k^2)$ as a function of longitudinal momentum fraction $x$ and transverse momentum $k$ reads 
\begin{multline}
{\cal F}(x,k^2) = {\cal F}^{(0)}(x,k^2)\;  + \\
+ \; \int_x^1 \frac{dz}{z} \int {dk'^2} \,\overline{\alpha}_s(k^2,k'^2) 
\bigg[ \frac{{\cal F}(\frac{x}{z},k'^2)}{|k^2-k'^2|}-\frac{k^2}{k'^2}\frac{{\cal F}(\frac{x}{z},k^2)}{|k^2-k'^2|}
+\frac{k^2}{k'^2}\frac{{\cal F}(\frac{x}{z},k^2)}{\sqrt{k^4+4k'^4}}
\bigg]
 + \\
 -\frac{2\pi^2}{N_c^2 R^2} \int_{x}^1\frac{dz}{z}\Bigg\{
\bigg[\int_{k^2}^{\infty} \frac{dl^2}{l^2} \, \overline{\alpha}_s(k^2,l^2) {\cal F}(\frac{x}{z},l^2)\bigg]^2 
+\\
+\; {\cal F}(\frac{x}{z},k^2)\int_{k^2}^{\infty} \frac{dl^2}{l^2} \, \overline{\alpha}^2_s(k^2,l^2)\ln\left(\frac{l^2}{k^2}\right) {\cal F}(\frac{x}{z},l^2)
\Bigg\}\; 
\; ,
    \label{eq:virtual_kms_F}
\end{multline}
where ${\cal F}^{(0)}$ is starting distribution. The linear part of the equation (the second line) corresponds to the angular averaged BFKL kernel whose real part represents emission of gluons and virtual corresponding to the reggeization. The nonlinear part (the third and fourth lines) accounts for recombination of gluons and is given by triple Pomeron vertex, here angular averaged \cite{Bartels:1994jj}. 
The strength of the nonlinear term is controlled by the effective   radius of proton or nucleus $R$,  which will be one of the fitting parameters.
In the coordinate space based derivation, this parameter originates from the assumption that the target is a large, homogeneous disk of radius 
$R$. After performing a Fourier transform from coordinate space to momentum space, integrating over the impact parameter, and normalizing the resulting gluon density so that $\int d^2b\;{\cal F}(x,k^2,b)={\cal F}(x,k^2)$ the nonlinear term gains the $1/R^2$ factor\footnote{For more details see recent review \cite{vanHameren:2023oiq}.}. 
In the above the rescaled coupling constant is $\overline{\alpha}_s(k^2,k'^2) = N_c\, \alpha_s\left(\mathrm{max}(k^2,k'^2)\right)/\pi$ with number of colors $N_c=3$.
In the LO the strong coupling is fixed in the high energy limit, it starts to run at NLO order.  The NLL calculation \cite{Fadin:1998py,Ciafaloni:1998gs} suggests that the natural scale as the argument of the running coupling is the virtuality of the emitted gluon $q^2=(k-k')^2$ (see \cref{fig: DIS F2}), see also a discussion in \cite{Ciafaloni:2003rd}. Here we use a choice of $\max(k'^2,k^2)$ which is numerically close to the above choice and also symmetric with respect to $k^2,k'^2$. In the present calculation we aim to solve \cref{eq:virtual_kms_F} down to very low transverse momenta. Therefore we need some form of regularization of the strong coupling. We shall take the parametrization \cite{Shirkov:1997wi,Kotikov:2019kci}
\begin{equation}
    \alpha_s(k^2) = \frac{12\pi}{33-2N_f}F(k,k_{fr},\Lambda_{QCD}) \; ,
\end{equation}
where 
\begin{equation}F(k,k_{fr},\Lambda_{QCD})=
    \left[\log\left(\frac{k^2+k^2_{fr}}{\Lambda_{QCD}^2}\right)\right]^{-1}-
    \frac{\Lambda_{QCD}^2}{k^2+k^2_{fr}-\Lambda_{QCD}^2} \;,
\end{equation}
with number of flavors $N_f=4$ parameter $k^2_{fr}=1.4 \, \rm GeV^2$  obtained by fitting the $\alpha_s$ from \cite{Lai:2010vv}. 
The formula follows from studies of analyticity properties of QCD coupling constant and was recently used in successful description of $F_2$ at small values of $Q^2$ \cite{Kotikov:2014faa}. Furthermore
it gives smooth extrapolation of the strong coupling to small transverse momenta.

In the following we shall  illustrate basic properties and behavior of the solution to \cref{eq:virtual_kms_F} at lowest order, including only the running coupling.

\begin{figure}[!t]
    \captionsetup{font={small,it}}  
    \centering
    \begin{minipage}{0.49\textwidth}
        \centering
        \includegraphics[width=\textwidth]{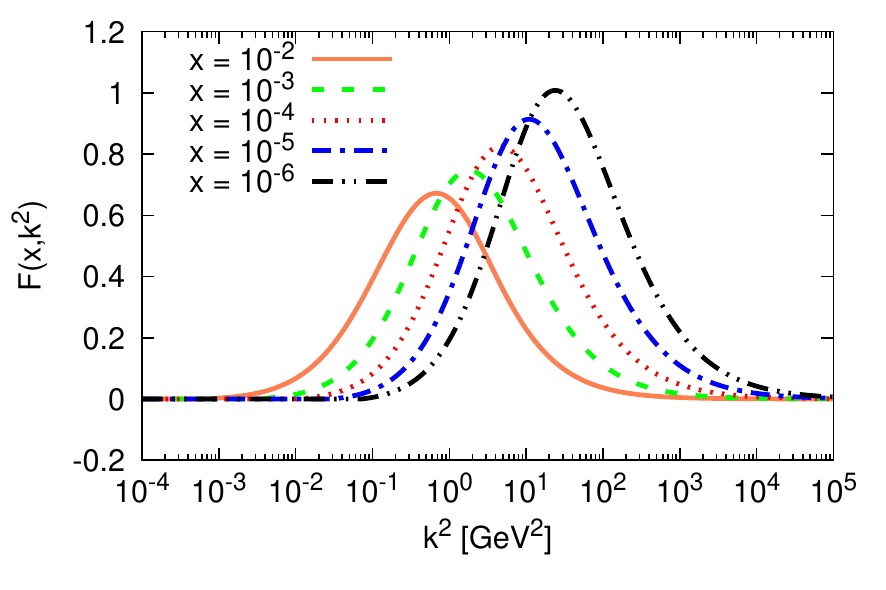}
        \caption{Solution of BK equation from \cref{eq:virtual_kms_F} as  function of $k^2$ for various values of $x$.}
        \label{fig:BKrc_k2}
    \end{minipage}
    \hfill
    \begin{minipage}{0.49\textwidth}
        \centering
        \includegraphics[width=\textwidth]{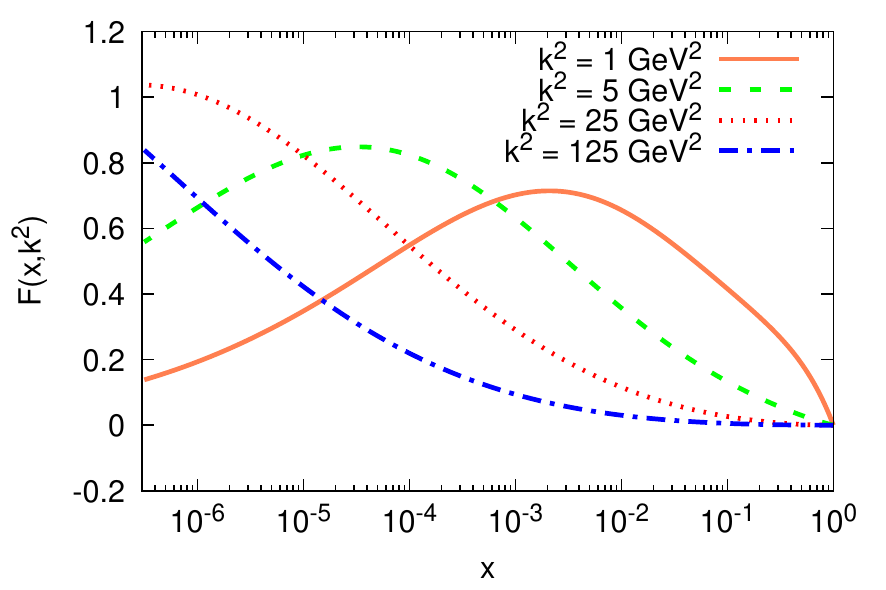}
       \caption{Solution of BK equation from \cref{eq:virtual_kms_F} as   function of $x$ for various values of $k^2$.}
        \label{fig:BKrc_x}
    \end{minipage}
\end{figure}

In \cref{fig:BKrc_k2,fig:BKrc_x} we present the solution of the  \cref{eq:virtual_kms_F} using the initial condition given by input motivated by the GBW model
\begin{equation}
{\cal F}^{(0)}(x,k^2)=(1-x)^{\alpha}k^2 e^{-k^2/Q_s^2} \; .
\end{equation}
In the above we set the saturation scale   $Q_s^2=1\; \rm GeV^2$  and $\alpha=1$, where we take $R^2 \approx 8.0\; \text{GeV}^{-2}$ in \cref{eq:virtual_kms_F}, which is the result of the our later reduced cross section fit in \cref{sec:fitresults}. The factor $(1-x)^{\alpha}$ proposed in \cite{Motyka:2002ww} regulates the behavior of the starting distribution at $x=1$. This is necessary especially in the case when DGLAP terms are included, as studied later in this section.   In \cref{fig:BKrc_k2} the solution is shown as a function of the transverse momentum for several values of $x$.  As $x$ becomes lower, the typical value of transverse momentum characterized by the saturation scale becomes larger and larger.  In \cref{fig:BKrc_x}, $x$ dependence is illustrated and we observe that,
for small transverse momentum, the gluon density has a valence-like behavior at small $x$. As $k^2$ becomes larger than the saturation scale, the falling part of gluon density happens at lower $x$.

\subsection{Higher order improvements to the BK evolution}

The equation discussed above is  LL in $\alpha_s\ln 1/x$ and  subject to large corrections \cite{Fadin:1998py,Ciafaloni:1998gs,Chirilli:2013kca} at higher orders. 
 Below we will use the prescription developed by \cite{Ciafaloni:2003ek,Ciafaloni:2003rd}.  
The  equation with resummed higher order corrections assumes the schematic form
\begin{equation}
    {\cal F}(x,k^2)= {\cal F}^{(0)}(x,k^2)+ {\cal K}_{res}\otimes{\cal F}(x,k^2)-{\cal V}\otimes  {\cal F}^2(x,k^2) \; ,
    \label{eq:schRBK}
\end{equation}
with 
\begin{equation}
{\cal K}_{res}\otimes {\cal F}\equiv \CK_0^{\rm kc}(z;k,k') \stackrel{z,k'}{\otimes} {\cal F} (\frac{x}{z},k')+\CK_0^{\rm coll}(z;k,k') \stackrel{z,k'}{\otimes} {\cal F} (\frac{x}{z},k') \; .
\label{eq:kterms}
\end{equation}
The first term in Eq.~\eqref{eq:kterms} is 
\begin{multline}
\CK_0^{\rm kc}(z;k,k') \stackrel{z,k'}{\otimes} {\cal F} (\frac{x}{z},k')= \\
 \; \int_x^1 \frac{dz}{z} \int {dk'^2} \,\overline{\alpha}_s(k^2,k'^2) 
\bigg[ \frac{{\cal F}(\frac{x}{z},k'^2)}{|k^2-k'^2|}\Theta\left(\frac{k^2}{z}-k'^2\right)-\frac{k^2}{k'^2}\frac{{\cal F}(\frac{x}{z},k^2)}{|k^2-k'^2|}
+\frac{k^2}{k'^2}\frac{{\cal F}(\frac{x}{z},k^2)}{\sqrt{k^4+4k'^4}} \, .
\bigg]
 \label{eq:LOBFKLkc}
\end{multline}

The kinematical (or consistency) constraint $\Theta\left(\frac{k^2}{z}-k'^2\right)$ term is implemented onto the real emissions only. It is here asymmetric, which corresponds to the asymmetric scale choice suitable for the DIS problem we are considering.
It is implemented as
\begin{equation}
k^{\prime 2} \le \frac{k^2}{z} \; .
\end{equation}

\begin{figure}[!t]
  \captionsetup{font={small,it}}
    \centering
    \begin{minipage}{0.49\textwidth}
        \centering
        \includegraphics[width=\textwidth]{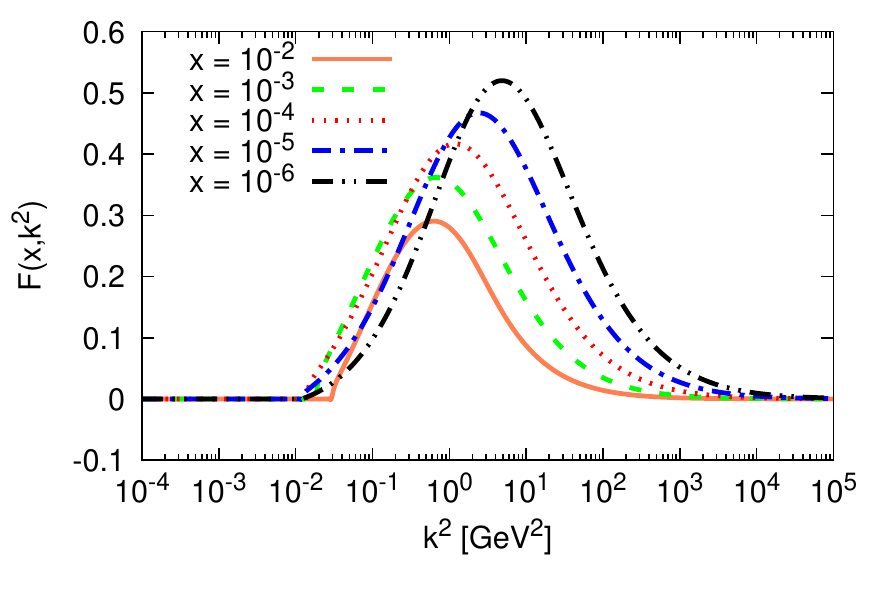}
        \caption{Solution of BK equation supplemented with kinematical constraint as a function of $k^2$ for various values of $x$.}
        \label{fig:kcBK1}
    \end{minipage}
    \hfill
    \begin{minipage}{0.49\textwidth}
        \centering
      \includegraphics[width=\textwidth]{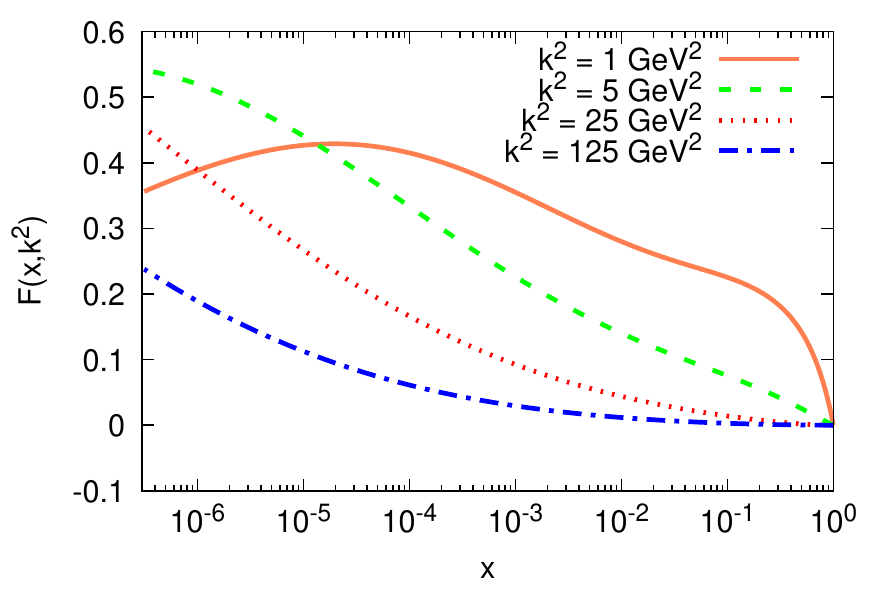}
        \caption{Solution of BK equation supplemented with kinematical constraint as a  function of $x$ for various values of $k^2$.}
        \label{fig:kcBK2}
    \end{minipage}
\end{figure}

In \cref{fig:kcBK1} we show solution of BK equation with kinematical constraint. We see that the normalization of gluon density changed almost by factor 2 as compared to the running coupling BK gluon in \cref{fig:BKrc_x,fig:BKrc_k2}. Growth with decreasing $x$ is slowed down and the valence like behavior is observed  for low values of $k^2$ only\footnote{For work addressing accounting for kinematical constraint in the coordinate space BK equation see \cite{Beuf:2014uia} pointing out relevance of kinematical constraint in the context of jet physics and NLO corrections see \cite{Caucal:2023fsf}.}.

\begin{figure}[!t]
  \captionsetup{font={small,it}}
    \centering
    \begin{minipage}{0.49\textwidth}
        \centering
        \includegraphics[width=\textwidth]{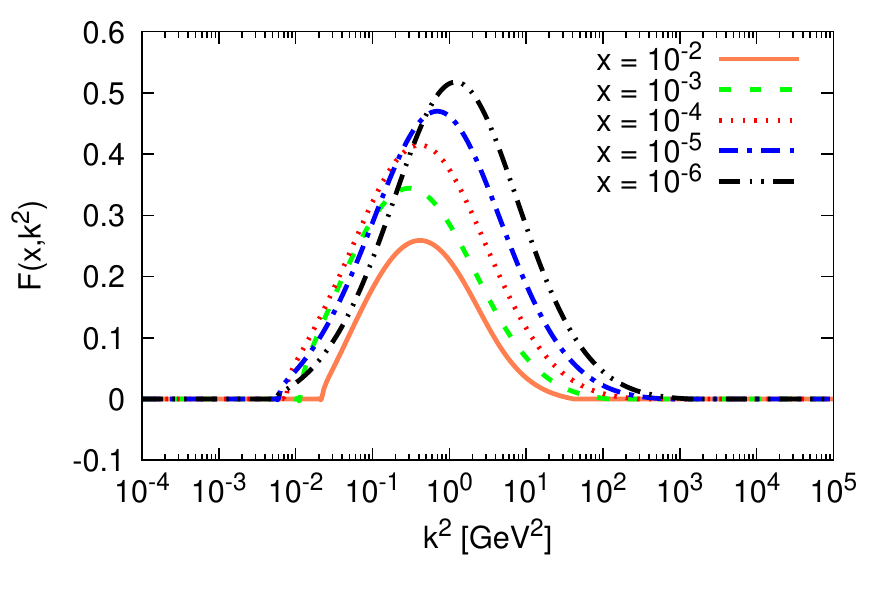}
        \caption{Solution of BK equation supplemented with kinematical constraint and DGLAP as a   function of $k^2$ for various values of $x$.}
        \label{fig:kcDGLAP1BK}
    \end{minipage}
    \hfill
    \begin{minipage}{0.49\textwidth}
        \centering
      \includegraphics[width=\textwidth]{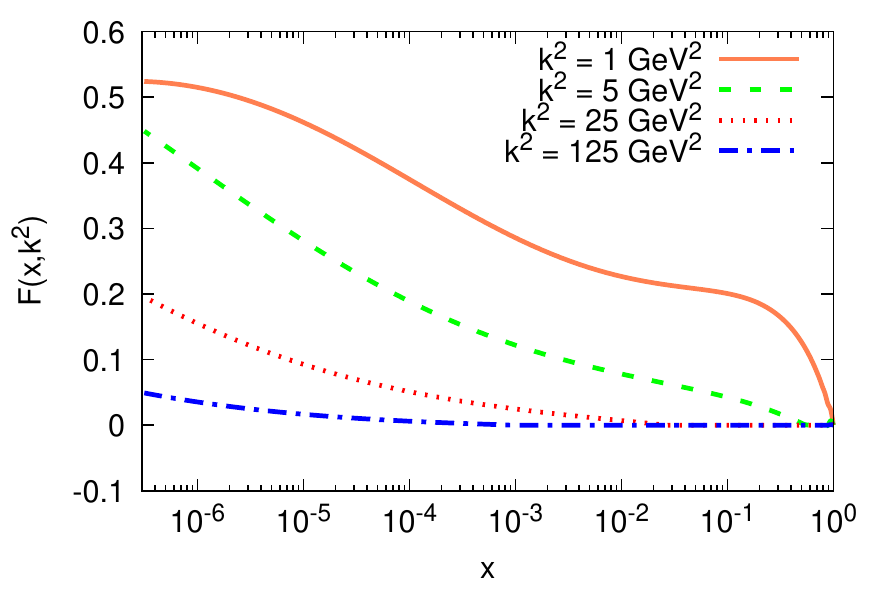}
        \caption{Solution of BK equation supplemented with kinematical constraint and DGLAP as a  function of $x$ for various values of $k^2$.}
        \label{fig:kcDGLAPBK2}
    \end{minipage}
\end{figure}

The second contribution in \eqref{eq:kterms}  is 
\begin{multline}
 \CK_0^{\rm coll}(z;k,k') \stackrel{z,k'}{\otimes} {\cal F}(\frac{x}{z},k')= 
  \int_x^1  \frac{dz}{z} \int_{0}^{k^2} \frac{{dk'}^2}{k^2} \;
  \bar{\alpha}_s(k^2) z\tilde{P}_{gg}(z)
  {\cal F}(\frac{x}{z},k')   \\
 + \int_x^1  \frac{dz}{z} \int_{k^2}^{k^2/z} \frac{{dk'}^2}{k'{}^2} \;
  \bar{\alpha}_s({k'}^2) z {\frac{{k^{\prime 2}}}{k^2}} \tilde{P}_{gg}(z {\frac{k^{\prime 2}}{k^2}})
  {\cal F}(\frac{x}{z},k') \;.  
  \label{eq:dglapterms}
\end{multline}
It is the sum of the collinear and anticollinear parts with the non-singular part of the splitting function
\begin{equation}
 \tilde{P}^{(0)}_{gg}  =  P^{(0)}_{gg} - \frac{1}{z} \;,
\end{equation}
where the $ P^{(0)}_{gg}$ is the   DGLAP gluon-gluon splitting function in LO given by
\begin{equation}
    P_{gg}^{(0)} = \frac{1-z}{z} + z(1-z)+\frac{z}{(1-z)_+}+\frac{1}{2C_A}\delta(1-z) \, \frac{11 C_A-4 N_f T_r}{6} \; .
\end{equation}
Here $C_A=3, T_r=1/2$ and $N_f$ is number of quark flavors. 
We stress that one of the big advantages of the formulation of the evolution in the momentum space is the ability to include the full DGLAP terms. 
In \cref{fig:kcDGLAP1BK} we present the solution of the BK supplemented with DGLAP correction and kinematical constraint. Compared to the scenario that accounts for running coupling and the kinematical constraint, the inclusion of DGLAP further suppresses the unintegrated gluon density, leading to a flatter $x$-spectrum.

\section{Structure function in the \(k_T\) factorization}
\label{sec:structurefunction}

\begin{figure}[!t]
    \centering
    \includegraphics[width=0.36\linewidth]{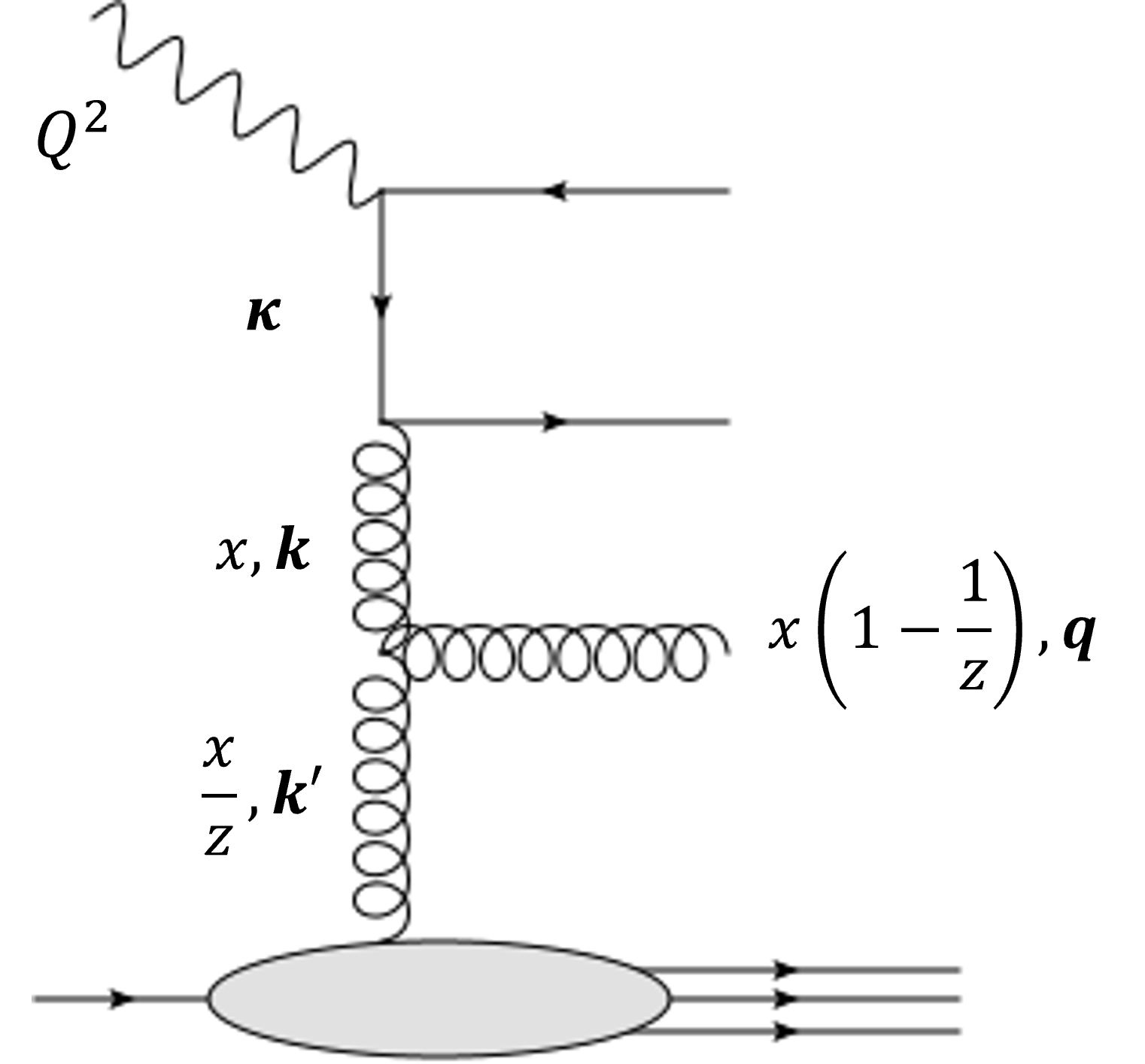}
    \caption{DIS diagram in the high energy factorization, where the transverse momentum of the emitting gluon $\bm{q} = \bm{k}' -\bm{k}$. }
    \label{fig: DIS F2}
\end{figure}

The DIS structure functions $F_2$ in the small $x$ region can be expressed as a convolution of the off-shell photon-gluon partonic cross section with the unintegrated gluon density. The longitudinal structure function $F_L$, which is a necessary contribution to compute the reduced cross section, can be calculated in a same fashion, see e.g. \cite{Golec-Biernat:2009mod}. This scheme, known as $k_T$ factorization \cite{Catani:1990eg,Catani:1990xk,Collins:1991ty}, is illustrated schematically in \cref{fig: DIS F2}. 

The perturbative contribution to the structure function \(F_2\) from the $k_T$ factorization is given by
\begin{equation}
	F_2(x,Q^2) = \sum_q \, e_q^2 \,  S_q(x,Q^2) \; ,
\end{equation}
where the sum is over the quark flavors and \( S_q(x,Q^2)\) can be factorized into the following form
\begin{equation}
	S_q (x, Q^2) \; = \; \int_x^1 \: \frac{dz}{z} \: \int
	\: dk^2 \: S_{\rm box}^q \: (z, m_q^2,k^2, Q^2) \: \mathcal{F} \left
	(
	\frac{x}{z}, k^2 \right ) \;.
	\label{eq:sqbox}
\end{equation}
Here, $S_{\rm box}^q$ is the off-shell photon-gluon partonic cross section (\textit{box} named after the calculation of the amplitude of \cref{fig: DIS F2} with its conjugate). 

The explicit expression of \cref{eq:sqbox} after the Sudakov decomposition is given by \cite{Askew:1992tw, Kwiecinski:1997ee}
\begin{align}
	\label{eq:boxc}
	S_q(x,Q^2)= &  \frac{Q^2}{4\pi^2}\int_{k_\text{min}^2} ^{\infty}\frac{dk^2}{k^2}\int_{0}^{1}d\beta\int d \kappa' \alpha_s(\mu^2) \left\{\left[\beta^2+(1-\beta^2)\right]\right. \nonumber \\ 
	& \left(\frac{\bm{\kappa}}{D_{1q}}-\frac{\bm{\kappa}-\bm{k}}{D_{2q}}\right)^2 + \left[m_q^2 + 4Q^2\beta^2 (1-\beta)^2\right]  \nonumber \\
	& \left.\left(\frac{1}{D_{1q}}-\frac{1}{D_{2q}}\right)^2\right\} \mathcal{F}\left(\frac{x}{z},k^2\right)\Theta\left(1-\frac{x}{z}\right),
\end{align}
where \( \bm{\kappa}\) and \(\bm{k}\) are quark and gluon transverse momenta respectively, and $\beta$ is the longitudinal momentum fraction of the photon carried by the quark variable defined in the  of the quark momentum (see Sudakov decomposition in \cite{Askew:1992tw}).
Instead of integrating on $\kappa$, it turns out to be convenient to work with the shifted quark transverse momentum \(\bm{\kappa}'=\bm{\kappa} -(1-\beta)\bm{k}\).  Masses of the quarks are denoted by $m_q$.

The energy denominators are defined as
\begin{eqnarray}
    	D_{1q} & = &\kappa^2+\beta(1-\beta)Q^2+m_q^2 \; ,\\
	D_{2q} & = &(\bm{\kappa}-\bm{k})^2+\beta(1-\beta)Q^2+m_q^2 \; .
\end{eqnarray}

In addition, the argument of the unintegrated gluon density from the resummed BK equation is $x/z$, where
\begin{equation}
    	z  = \left[1+\frac{\kappa'^2+m_q^2}{\beta (1-\beta) Q^2}+\frac{k^2}{Q^2}\right]^{-1} \; .
\end{equation}

This prescription  above arises  from a rigorous treatment of the photon-gluon fusion process, incorporating its exact kinematic \cite{Askew:1992tw}. In the $k_T$ factorization framework, such a precise kinematic description elevates the photon impact factor beyond LL approximations \cite{Bialas:2000xs,Bialas:2001ks} yielding significant phenomenological consequences, as demonstrated in \cite{Golec-Biernat:2009mod}. Notably, recent advances have enabled collinear resummation to the photon impact factor \cite{Colferai:2023dcf}, which retain consistency with the exact  kinematic formulation in the collinear limit\footnote{For recent work on evaluation of $F_2$ in coordinate space at NLO accuracy see \cite{Beuf:2020dxl}. }.

We take the argument of the strong coupling \(\alpha_s\) to be \(\mu^2=k^2+\kappa'^2+m_q^2\), while the masses of quarks are \(m_u=m_d=m_s=0\) and \(m_c=1.4\ \text{GeV}\).

In principle, the $k_T$ factorization approach requires integrating over the transverse momentum $\kappa$ down to zero in \cref{eq:boxc}, which extends into the non-perturbative region.
Meanwhile, although the non-linear BK evolution allows the unintegrated gluon density $\mathcal{F}(x, k^2)$ to be evolved into the low gluon transverse momentum $k$ region, it does not alter the infrared behavior of the quark transverse momentum $\kappa$. Therefore following \cite{Kwiecinski:1997ee}, we include a separate non-perturbative contribution. 
This follows from the well known fact, the  structure function $F_2$ receives large soft contribution. As proposed in  \cite{Askew:1993jk}, this soft or non-perturbative contribution has been simply parametrized as the constant background term in addition to the perturbative contribution.  In the approaches within the dipole model, the non-perturbative contribution corresponds to  integration over the large dipole sizes with the flat dipole cross section, see for example discussion in \cite{Berger:2011ew,Mantysaari:2018zdd}.  
For small values of $\kappa$, this non-perturbative region corresponds to the so-called \textit{soft Pomeron exchange} contribution \cite{Donnachie:1992ny}, which can be parametrized phenomenologically as
\begin{equation}
	S^{(a)} \; = \; S_u^\funp \: + \: S_d^\funp \: + \: S_s^\funp \; ,
	\label{eq:soft0}
\end{equation}
for light quarks and
\begin{equation}
		S_u^\funp \; = \; S_d^\funp \; = \; 2S_s^\funp \; = \; C_\funp \:
	x^{-\lambda} \: (1 - x)^8,
	\label{eq:soft}
\end{equation}
where coefficient \( C_\funp\) is a free parameter independent of \(Q^2\) and $\lambda$ is the soft Pomeron power to be fitted from experimental data. We assume that the charm contribution is generated purely perturbatively, and therefore there is no soft Pomeron contribution corresponding to it, i.e. $S_c^\funp = 0$.

By noting the perturbative contribution to the structure function in \cref{eq:boxc} as $S^{\text{pert}}$, the total structure functions $F_2$ is therefore simply
\begin{equation}
    F_2 = \sum_q e_q^2 \left( S^\funp_q + S^{\text{pert}} \right).
    \label{eq:F2}
\end{equation}

\section{Fit results}
\label{sec:fitresults}

In this section, we present the fit to HERA data \cite{H1:2015ubc} and the corresponding  unintegrated gluon density, i.e. the solution  of \cref{eq:schRBK} . 
We choose the following initial condition inspired by the GBW model,
\begin{equation}
    {\cal F}^{(0)}(x,k) \;=A\,\alpha_s(k^2)(1-x)^{\alpha}x^{\beta}
   {(k^2)^{\gamma}}e^{-B^2 k^2}
.
    \label{eq:GBWmod_gammas}
\end{equation}

The parameters are fitted to the reduced cross section $\sigma_r$, which is given by the combination of the two structure functions $F_2$ and $F_L$
\begin{equation}
\sigma_r(y,x,Q^2)=F_2(x,Q^2)-\frac{y^2}{1+(1-y)^2}F_L(x,Q^2).
\end{equation}
Here, the inelasticity $y=Q^2/(s x)$ and $\sqrt{s}$ is the electron-proton collision energy. We set the the flavor number $N_f = 4$.

\begin{figure}[t!]
  \captionsetup{font={small,it}}
        \centering
        \includegraphics[width=\textwidth]{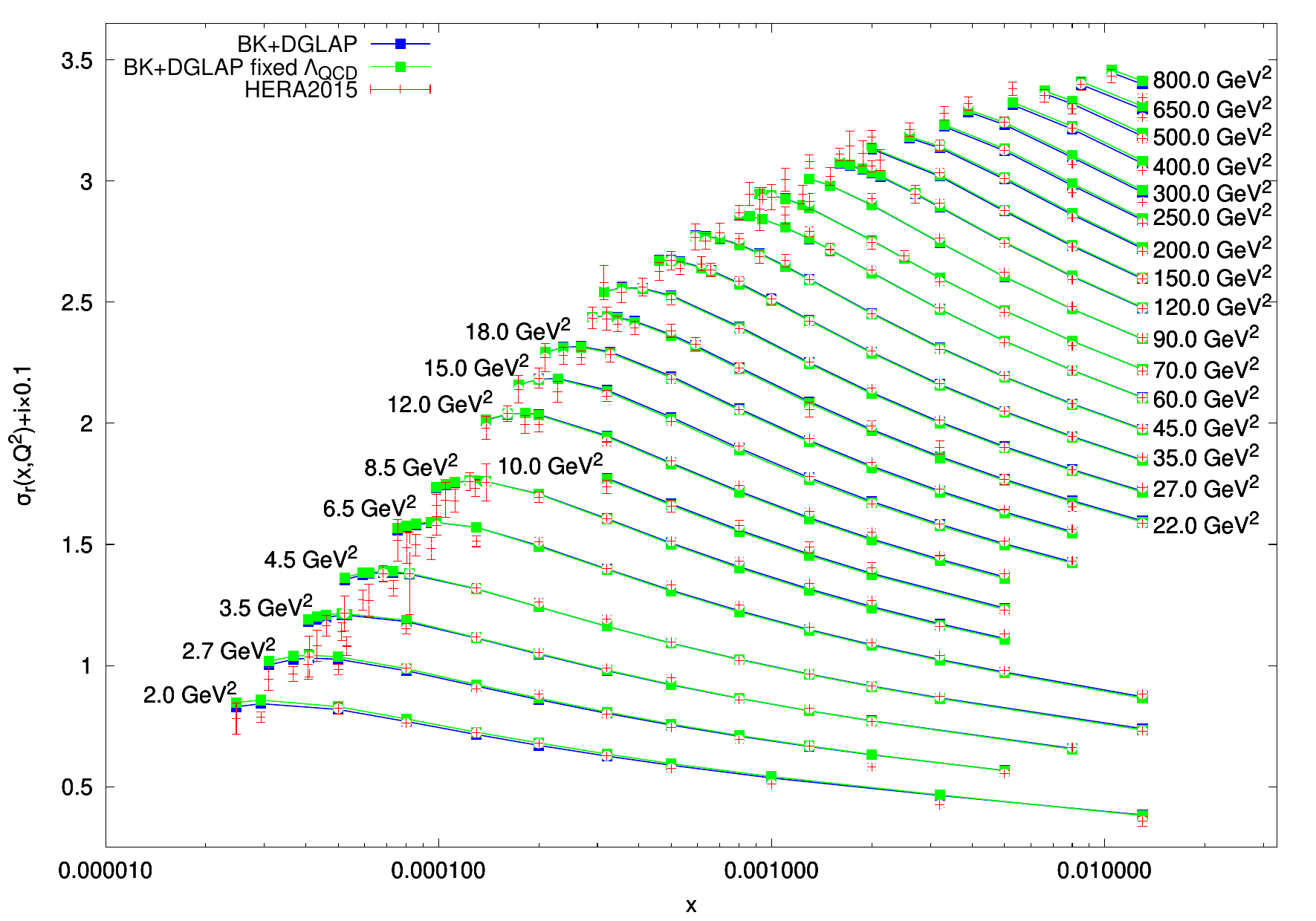}
        \caption{Comparison between our reduced cross section fit with the HERA data. Each curve is separated with an additional offset 0.1 to better present a wider $Q^2$ spectrum. The blue curve is for fitted $\Lambda_{\rm QCD}$ and green curve for fixed $\Lambda_{{\rm QCD}} = 0.289\; {\rm GeV} $.}
        \label{fig:sigred}
\end{figure}

In \cref{fig:sigred} we present our fit compared with HERA data. By selecting all low x data ($x < 0.013$), we fit a total of 239 data points. We propose two fit scenarios: one with fixed $\Lambda_{\text{QCD}} = 0.289\; \text{GeV}$, matching the choice in \cite{Li:2022avs}, and the other where  $\Lambda_{\text{QCD}}$ is treated as a free fit parameter. Both fits provide a consistent description of the HERA data, and the obtained fit parameter values are summarized in \cref{tab:combined}.

\begin{table}[h]
\centering
\resizebox{\textwidth}{!}{%
\begin{tabular}{l|l|l|l|l|l|l|l|l|l}
\hline
 & $\alpha$ & $\beta$ & $A$ & $B^2$ & $\gamma$ & $C_\funp$ & $\lambda$ & $R^2$ & $\Lambda_{\rm{QCD}}$ \\
\hline
$\chi^2/\mathrm{dof}=1.6$
 & 0.80662 & -0.42651 & 0.74038 & 0.64239 & 1.09987 & 0.6163 & -0.02361 & 2.50003 & 0.499 \\
$\chi^2/\mathrm{dof}=2.0$
 & 0.67356 & -0.41625 & 0.65917 & 0.47992 & 1.05338 & 0.572 & -0.00223 & 2.54913 & fixed \\
\hline
\end{tabular}%
}
\caption{$\chi^2$ and fit parameters in the fitted  and fixed $\Lambda_{\text{QCD}}$ scenarios respectively. }
\label{tab:combined}
\end{table}

\begin{figure}[t!]
  \captionsetup{font={small,it}}
    \centering
    \begin{minipage}{0.49\textwidth}
        \centering
        \includegraphics[width=\textwidth]{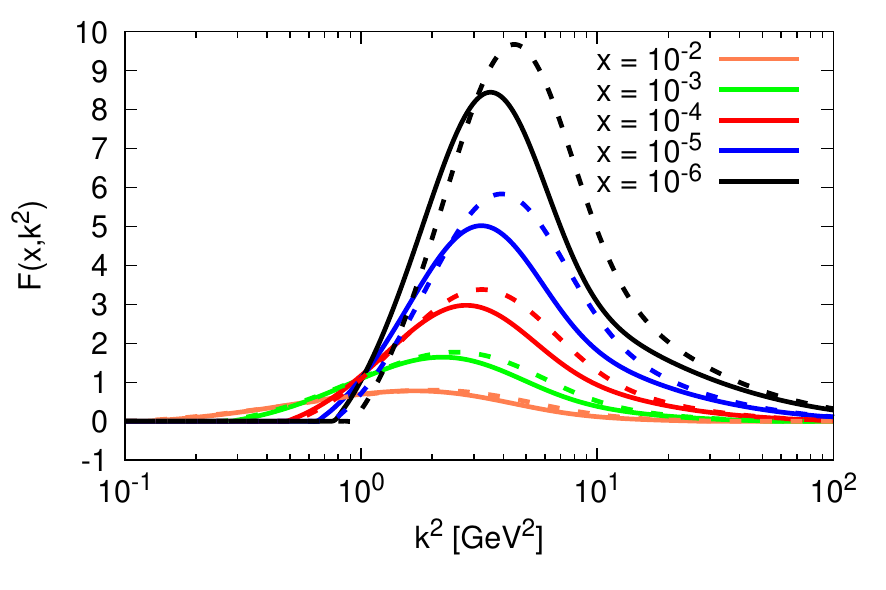}
        \caption{Plots of fitted unintegrated gluon density as a function of $k^2$ for various values of $x$. Solid and dashed curves denote fitted and fixed $\Lambda_{\rm{QCD}}$.}
        \label{fig:fitglue1}
    \end{minipage}
    \hfill
    \begin{minipage}{0.49\textwidth}
        \centering
      \includegraphics[width=\textwidth]{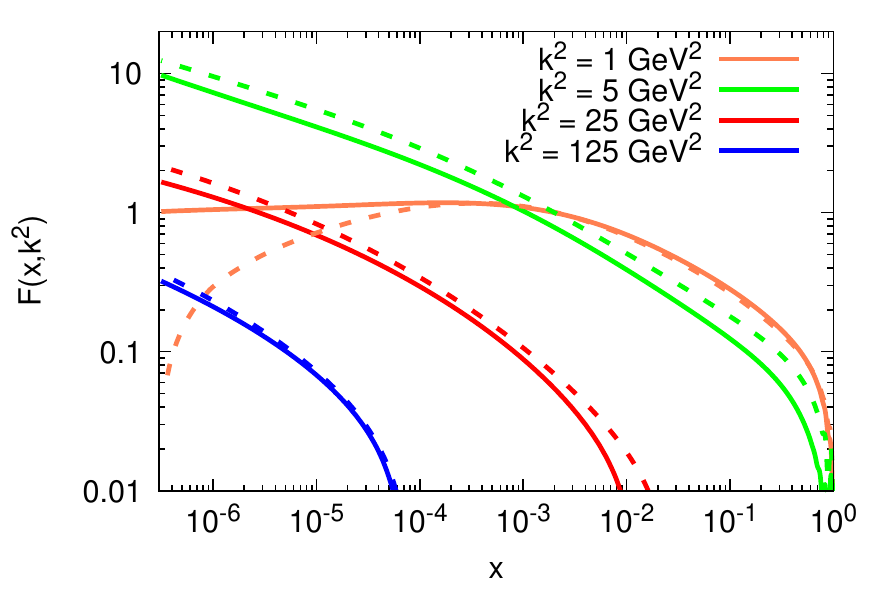}
        \caption{Plots of fitted unintegrated gluon density as a function of $x$ for various values of $k^2$. Solid and dashed curves denote fitted and fixed $\Lambda_{\rm{QCD}}$.}
        \label{fig:fitglue2}
    \end{minipage}
\end{figure}

The solutions with fitted $\Lambda_{\text{QCD}}$ (solid curve) and fixed $\Lambda_{\text{QCD}}$ (dashed curve) are presented in \cref{fig:fitglue1} and \cref{fig:fitglue2}. The cusp appearing for lowest values of $x$ around $k^2 \sim 1 \; \text{GeV}^2$ in \cref{fig:fitglue1} is due to the cut we enforced in the solver to prevent slightly negative values of the unintegrated gluon density. In \cref{fig:fitglue1}, both transverse momentum spectrum of the fitted and fixed $\Lambda_{\text{QCD}}$ fit resembles very much the  the characteristic scaling behavior of the saturation scales, i.e. as $x$ becomes smaller the saturation scale grows and gluon momenta are pushed toward perturbative region.

In \cref{fig:fitglue2}, we observe that the gluon densities in two scenarios are generally close, except for the curves with  $k^2 = 1\; \text{GeV}^2$, where our negativity limiter has a more significant impact.  We comment that the underlying negative contributions stem from the interplay of negative nonlinear term in the BK equation and the collinear resummations in the linear part of the equation. Notably, our current resummed BK equation incorporates the kinematical constraint only in the linear term, making negative values more likely to appear. A possible improvement would be to introduce the kinematical constraint also in the nonlinear term. At present, such a constraint on the nonlinear term has been implemented in the BK equation only within the dipole model and in position space \cite{Beuf:2014uia,Watanabe:2015tja,Albacete:2015xza}. We plan to further investigate its effects within our formalism in future work.

\begin{figure}[t!]
  \captionsetup{font={small,it}}
    \centering
    \begin{minipage}{0.49\textwidth}
        \centering
        \includegraphics[width=\textwidth]{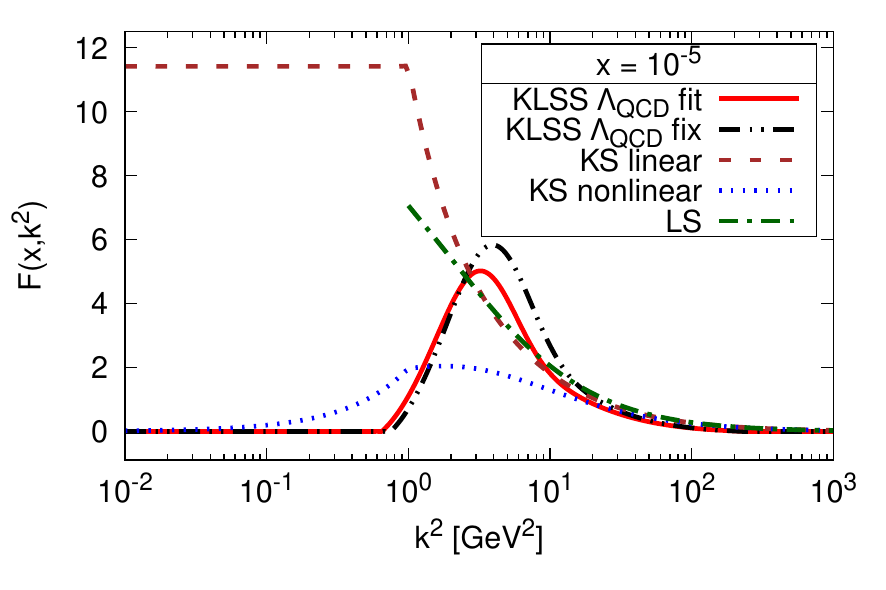}
        \caption{Plots of fitted unintegrated gluon density (with fixed and fitted $\Lambda_{\rm{QCD}}$) as a function of $k^2$ for $x=10^{-5}$ compared to KSlinear, KS nonlinear and LS linear gluon densities.
        }
        \label{fig:fitglue3}
    \end{minipage}
    \hfill
    \begin{minipage}{0.49\textwidth}
        \centering
      \includegraphics[width=\textwidth]{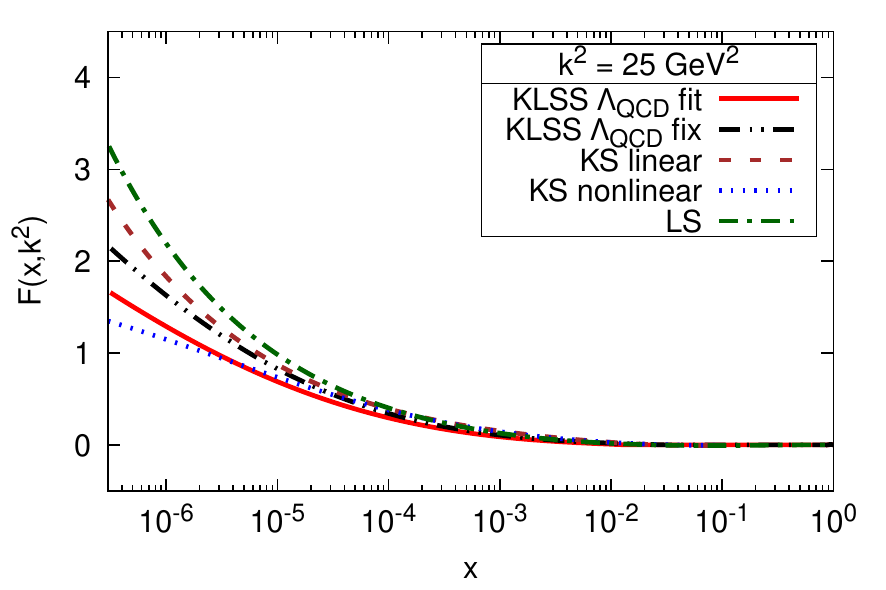}
        \caption{Plots of the unintegrated gluon densities (with fixed and fitted $\Lambda_{\rm{QCD}}$) as a function of $x$ for $k^2=25 \;{\rm GeV}^2$ compared to KSlinear, KS nonlinear and LS linear gluon densities.}
        \label{fig:fitglue4}
    \end{minipage}
\end{figure}

In \cref{fig:fitglue3}, \cref{fig:fitglue4} we show comparison of obtained gluon density to KSlinear, KSnonlinear \cite{Kutak:2012rf} and Li-Stasto(LS)\cite{Li:2022avs}. The main difference between  our gluon density and the other  distributions is that we account for perturbative evolution down to low $k$ of gluon. In practice, we set the lower bound  $k_{\text{min}}^2 = 10^{-6} \; \text{GeV}^2$. This is possible because the nonlinear BK term regulates the infrared behavior of the linear BFKL evolution.  The consequence of that is that the integral with the box \cref{eq:sqbox} can be  evaluated directly with \cref{eq:boxc} without splitting it to perturbative and non-perturbative parts, where a separated collinear approximation is typically applied \cite{Li:2022avs,Kutak:2012rf,Kwiecinski:1996td}. This leads to larger gluon density as compared to other gluon densities. We note that in the KS nonlinear, the method of solving the equation did not allow to go to transverse momentum smaller than 1\;GeV$^2$ and extrapolation was used in this region. The $F_2$ was calculated in similar manner as in KS linear and LS. 

\section{Conclusions}
\label{sec:conclusions}

In this paper we investigated the momentum space  BK equation with running coupling, kinematical constraint and  DGLAP  terms both in collinear and anticollinear limit.
We  first studied  solutions for particular cases: including only the  effects of kinematical constraint and kinematical constraint together with DGLAP terms. We demonstrated that each of these contributions have substantial impact on resulting solution and as it changes overall change normalisation and shape of gluon density.
This is in agreement with previous findings by \cite{Kutak:2004ym}. However the present formulation allows for extending the solution to small transverse momentum scales.  This procedure impacts the resulting gluon density in  such way that in non saturated region it is larger than the KS \cite{Kutak:2012rf} gluon density while it is smaller in the saturated region. This may have important phenomenological consequences e.g for jet physics \cite{Marquet:2007vb,Albacete:2014fwa, vanHameren:2016ftb,Deak:2009xt,Deak:2010gk,Kutak:2012rf,vanHameren:2023oiq}. 

We used the obtained dipole gluon density together with impact factor with exact kinematics to calculate DIS reduced cross section. The form of the starting distribution was motivated by the Golec-Biernat-Wuesthoff model, but included additional  modifications and parameters as required by the fit. 
Finally we would like to add that although  there is some modeling involved the equation that we proposed is in line of the CCSS resummation, in order  to complete it one needs to account for remaining NLO contributions as it has been done in the linear case in \cite{Li:2022avs}. We leave the study of  equation with these modifications  for the future research.

\section*{Acknowledgments}
KK thanks Hannes Jung for informative discussions. 
KK acknowledges support from grant No. 2019/33/B/ST2/02588 (2020-23). AMS is  supported by the U.S. Department of Energy grant No. DE-SC-0002145 and within the framework of the of the Saturated Glue (SURGE) Topical Theory Collaboration. WL is supported by the National Science Foundations of China under Grant No. 12275052, No. 12147101. RS is supported by the AGH University of Science and Technology, Faculty of Metals Engineering and Industrial Computer Science, Work no. 16.16.110.663.

\bibliographystyle{JHEP} 
\bibliography{references}

\end{document}